# Spatial variation of energy gap and Landau levels around gapped bilayer graphene domain walls

Long-Jing Yin[1,2], Si-Yu Li[1,2], Jia-Bin Qiao[1,2], Wei-Jie Zuo[1,2], and Lin He[1,2],*

[1] Department of Physics, Beijing Normal University, Beijing, 100875, People's Republic of China

[2] Center for Advanced Quantum Studies, Beijing Normal University, Beijing, 100875, People's Republic of China

* Email: helin@bnu.edu.cn

**Bilayer graphene contains, compared to graphene monolayer, an additional graphene sheet and, therefore, extra degrees of freedom, making it a unique system for complex electronic states to emerge. Here, we show that there are two types of domain walls, *i.e.*, coupling domain wall and potential domain wall, in gapped graphene bilayers, which make microscopic electronic properties of the bilayers varying spatially. The coupling domain wall separates two graphene bilayer regions with different interlayer coupling strengths and the potential domain wall is a boundary separating two adjacent regions with different chemical potentials between two layers. We present a microscopically study, using scanning tunnelling microscopy and spectroscopy, around the two types of domain walls. The well-defined domain walls allow us to spatially resolve the energy gap and Landau levels around them, which show novel behaviour beyond what is expected from current theoretical models. Our result indicates that the graphene bilayer may exhibit exotic electronic properties related to the microscopic symmetry of the two layers.**

Gapped bilayer graphene has attracted much attention not only because its application in nanoelectronic and nanophotonic devices[1-7], but also because the valley-related electronic properties and novel electronic states[8-17]. Recently, it was predicted that one-dimensional (1D) chiral boundary states of quantum valley Hall insulators can be realized in graphene bilayer by using smooth domain walls[18-22]. Both the domain walls between AB- and BA-stacked bilayer graphene and the domain walls between band-inverted gapped bilayer graphene generated by opposite vertical electrical fields are predicted to host the 1D symmetry-protected topological states. Very recently, the topologically protected 1D chiral states have been observed along the AB-BA domain walls through transport menasurement[23], paving the road to explore unique electronic states in graphene bilayer.

Here, we use scanning tunnelling microscopy (STM) and spectroscopy (STS) measurements to study the microscopic electronic properties of gapped graphene bilayers, and show that smooth domain walls separating two adjacent bilayer regions with different electronic properties are widespread. In the quantum Hall regime, we demonstrate that the energy gap and Landau levels vary spatially around the domain walls.

The low-energy bilayer Hamiltonian with nearest-neighbor intralayer $t$ and interlayer hopping $t_\perp$ can be written as[18-22]

$$\begin{aligned} \boldsymbol{H} = &-t\sum_{i=1}^{2}\sum_{m,n} a_{i,m,n}^{\dagger}(b_{i,m,n} + b_{i,m-1,n} + b_{i,m,n-1}) - t_\perp \sum_{m,n} a_{1,m,n}^{\dagger} b_{2,m,n} \\ &+ U\sum_{i=1}^{2}\sum_{m,n}(-1)^{i}(a_{i,m,n}^{\dagger}a_{i,m,n} + b_{i,m,n}^{\dagger}b_{i,m,n}) + H.c. \end{aligned} \quad (1)$$

where $i = 1, 2$ is a layer index, $U$ is the chemical potential difference between two layers, and the operators $a^\dagger_{i,m,n}$ and $b^\dagger_{i,m,n}$ ($a_{i,m,n}$ and $b_{i,m,n}$) create (annihilate) an electron at the two-site unit cell (*m*,*n*) of layer *i*. In the Hamiltonian (1), the first term describes the nearest-neighbor intralayer hopping on the honeycomb lattice, the second term represents the nearest-neighbor interlayer hopping of AB-stacking bilayer, and a nonzero *U* of the third term could induce a finite gap in the graphene bilayer. Spatial variations of the interlayer hopping $t_\perp$ and the chemical potential difference *U* could induce two types of domain walls, *i.e.*, coupling domain wall and potential domain wall, as schematically shown in Fig. 1. We study systematically the electronic properties of the two types of domain walls in the graphene bilayers.

The STM and STS measurements were carried on Highly Oriented Pyrolytic Graphite (HOPG) surface (see Supplementary Information for details). The topmost few layers of the HOPG usually decouples from the bulk[24-28] and, very importantly, it is convenient to identify decoupled graphene single-layer, Bernal (AB-stacking) bilayers, and Bernal trilayers on the HOPG substrate through their tunnelling spectra in high magnetic fields[27]. In our experiment, we first identify the decoupled Bernal bilayers on HOPG surface by high fields STS and only these decoupled bilayer regions are further characterized in detail in this work.

Figure 2a shows a representative STM image of a decoupled bilayer region around a coupling domain wall. Comparing to that on the right side of the domain wall, the interlayer distance on the left side of the domain wall is slightly enlarged ~ (25 ± 5) pm, as shown in Fig. 2b, which indicates that there are different interlayer coupling strengths in the adjacent two regions. Our STS measurements, as shown in Fig. 2c-2f, really demonstrate that the two adjacent regions separating by the domain wall exhibit quite different electronic properties. On the left side, ~ 10 nm away from the domain wall, the zero-field spectrum displays a V-shaped tunneling conductance (Fig. 2c). In the presence of high magnetic fields, the spectra exhibit Landau quantization of Dirac fermions (Fig. 2c), which is identical to that observed in a gapped graphene monolayer[24,29-31]. The observed Landau-level (LL) energies in a gapped graphene monolayer should be described by[31,32]

$$E_n = \text{sgn}(n)\sqrt{(\hbar\omega_B)^2|n|+\Delta^2} + E_0, \qquad n = ...-2,\ -1,\ 0,\ 1,\ 2..., \qquad (2)$$

where $\hbar\omega_B = \sqrt{2}\hbar v_F/l_B$, $l_B = \sqrt{c\hbar/(eB)}$ is magnetic length, $e$ is the electron charge, $v_F$ is the Fermi velocity, $E_0$ is the energy of charge neutrality point, and $2\Delta$ is the site energy difference between the A and B sublattices. The fitting of the experimental data to Eq. (2), as shown in Fig. 2d, yields $v_F = (0.82 \pm 0.01) \times 10^6$ m/s and the energy gap $2\Delta = 18$ meV. Two important results can be concluded from the STM and STS measurements. First, a slightly enlarged interlayer distance, ~ (25 ± 5) pm, could efficiently decouple the adjacent bilayers, *i.e.*, the interlayer hopping $t_\perp$ is almost reduced to zero in the left side of the domain wall. This agrees well with previous studies that graphene bilayers will exhibit electronic properties of graphene monolayer when the interlayer distance is slightly larger than 0.34 nm[24,27,30]. Second, the electrostatic potential of the second commensurate layer still can break the inversion symmetry of the decoupled topmost graphene sheet[30], as shown in the triangular contrasting of the STM image in Fig. 2b, and generate a finite gap in it. In the quantum Hall regime, the energies of the $n = 0$ LL in the $K$ and $K'$ valleys are shifted in opposite directions by the inversion symmetry breaking and, therefore, the $n = 0$ LL is split into the $0_-$ and $0_+$ LLs (here +, - are valley indices)[31,32], as shown in Fig. 2c. We further demonstrate that the amplitude of the wavefunction of the $0_-$ LL is mainly on one of the sublattice and that of the $0_+$ LL is mainly on the other sublattice of graphene (see Supplementary Information Fig. S1 for details), characterizing the internal structure of the two-component spinor of the Dirac-Landau level[31].

On the right side, ~ 10 nm away from the domain wall, the spectra recorded both in zero magnetic field and high magnetic fields (Fig. 2e) exhibit characteristics that are expected to be observed in gapped graphene bilayers[8,27,33]. The substrate (here the graphite) breaks the symmetry of the topmost adjacent bilayers and generate a finite gap in the parabolic bands of the Bernal bilayer. In zero magnetic field, there is a pronounced peak located at about 30 mV in the spectrum (Fig. 2e), which is attributed to the density of states (DOS) peak generated at the conduction-band edge (CBE) of the gapped bilayer. In the quantum Hall regime, the LL energies of a gapped graphene bilayer follow[8,27,33]

$$E_n = \pm [(\hbar\omega_c)^2(n(n-1)) + (U/2)]^{1/2} - \xi z U/4, \quad n = 2,3,4\ldots$$

$$E_0 = \xi U/2, \qquad E_1 = \xi (U/2)(1-z), \qquad (3)$$

where $\omega_c = eB/m^*$ is the cyclotron frequency, $m^*$ is the effective mass of quasiparticles, and $\xi = \pm$ represents for the valley indices. For $B \leq 8$ T, $z = 2\hbar\omega_c / t_\perp \ll 1$, and $|U|$ is approximately equivalent to the energy gap when $U < t_\perp$. According to the fitting result shown in Fig. 2f, the band gap $E_g$ ($\approx U$) of the bilayer region on the right side of the domain wall is estimated to be about 20 meV and the effective mass is $m^* = (0.034 \pm 0.004)m_e$ ($m_e$ is the free-electron mass). Both the gap and the effective mass are consistent well with the range of values reported previously for Bernal bilayers[2-10,27]. Here, we should point out that the valley-polarized quartet $LL_{(0,1,+)}$, which is generated at the CBE, is mainly localized on the topmost graphene layer. The other quartet $LL_{(0,1,-)}$, which is generated at valence band edge, resides mainly on the second layer. The STS spectra probe predominantly the DOS of the top layer, therefore, the signal of the quartet $LL_{(0,1,+)}$ is much stronger in the spectra (Fig. 2e).

The result in Fig. 2 demonstrates that the coupling domain wall is a boundary between the single-layer-like region (the left side) and the bilayer-like region (the right side). We now show STS spectra recorded across the domain wall, as shown in Fig. 3. With increasing the interlayer hopping (or with decreasing the interlayer distances) from the left region to the right region, we observe evolution of the zero-field spectra across the domain wall: the spectra change gradually from the single-layer feature to the bilayer characteristic (Fig. 3a). In the quantum Hall regime, the evolution from the Landau quantization of single-layer graphene in the left region to that of bilayer graphene in the right region is even more pronounced (Fig. 3b). On the left and right sides away from the domain wall, the recorded spectra show well-defined Landau quantization of single-layer graphene and bilayer graphene, respectively. In the domain wall, the LLs vary spatially and connect different levels of the single-layer and bilayer graphene in the opposite side, as shown in Fig. 3b. Therefore, these domain wall LLs cannot be described by neither Eq. (2) nor Eq. (3).

Here we should point out that the observed LLs in the domain wall also differ quite from the interface LLs predicted to be observed in a graphene monolayer-bilayer planar junction with atomically sharp boundary[34], which is predicted to lift the valley degeneracy of the LLs in the opposite side. In our experiment, we do not observe such a valley splitting of the LLs around the domain wall and the recorded LLs evolve gradually from Landau quantization of the graphene monolayer to that of the Bernal bilayer across the domain well, which is about 10 nm in width.

The disorder potentials induced by substrates[35-37] could strongly affect the spatial variation of the potential difference between the adjacent two layers[8] [the third term of Hamiltonian (1)] and, therefore, divide graphene bilayers into different regions with various energy gaps. For the special case that the band gaps of the two adjacent bilayer regions are inverted, the potential domain wall between them is predicted to host the 1D symmetry-protected topological states[18-22]. Generally, the two adjacent bilayers observed in our experiment just have different values of the band-gap, as shown in Fig. 4. Figure 4a shows a representative STM image of a decoupled Bernal bilayer separating by a potential domain wall. The decoupling behavior of the topmost bilayer is attributed to a large twisted angle, ~ $\theta = (6.2 \pm 0.1)°$, between the second graphene layer and the third layer[26,38-40]. The fixed-bias conductance map, as shown in Fig. 4b, shows the spatial distribution of the disorder potential around the domain wall of the bilayer. The tunnelling spectra recorded in opposite side of the domain wall under different magnetic fields, as shown in Fig. 4c and 4e, show high quality Landau quantization of massive Dirac fermions, and the LL peaks are even well resolved up to $n = 10$ (see Supplementary Information Fig. S2). The Landau quantization on both positions C1 and C2 can be described quite well by Eq. (3), as shown in Fig. 4d and 4f. On position C1, the energy gap $E_g = 32$ meV and the effective mass of quasiparticles $m^* = (0.036 \pm 0.002)m_e$. On position C2, we obtain $E_g = 60$ meV and $m^* = (0.033 \pm 0.003)m_e$. Therefore, the potential domain wall (Fig. 4a and 4b) is a boundary separating the two graphene bilayer regions with the gaps of 32 meV and 60 meV respectively (as schematically shown in Fig. 5a).

To investigate the electronic properties of the potential domain wall, we systematically measured the spatial-resolved STS spectra across it, as shown in Fig. 5. Figure 5b shows several representative spectra recorded at different positions across the domain wall in a magnetic field of 8 T, which indicate that the charge neutrality point $N_C$ increases gradually from about 21 meV at the position C1 to about 43 meV at the position C2. The conductance map shown in Fig. 5c clearly reveals an increase of the band gap from the position C1 to the position C2. More importantly, we plot d$I$/d$V$ spectra at different magnetic fields as a function of positions across the potential domain wall in Fig. 5d-5f. In the quantum Hall regime, the LLs vary spatially in the domain wall and connect different levels of the bilayer regions in the opposite side, as shown in Fig. 5e and 5f. A close examination of these LLs maps in the domain wall reveals unexpected behavior that is beyond the expectation of current theoretical models. Below the $N_C$, the LLs with the same indices in the opposite side of the domain wall connect. However, above the $N_C$, the LL$^2_{(0,1,+)}$ (the superscript 1/2 marked the LLs belong to the C1/C2 region) in the C2 region connects the 2$_{th}$ LL in the C1 region, and the $n_{th}$ LL ($n \geq 2$) in the C2 region connects the $(n+1)_{th}$ LL in the C1 region, as shown in Fig. 5e and 5f. As a result, the $LL^1_{(0,1,+)}$ in the C1 region is left to be "unpaired" and it extends to quite a long distance in the C2 region (see Supplementary Information Fig. S3 for details). This unexpected result allows us to detect the signal of $LL^1_{(0,1,+)}$ in the C2 region with a distance of more than 10 nm away from the domain wall. In zero magnetic field, we can even detect the signal of the CBE of the C1 region in the C2 region with a distance of about 50 nm away from the domain wall (Fig. 5d), suggesting an unexplored proximity effect between the adjacent gapped graphene bilayers. This result indicates that the bilayer regions with smaller gaps may affect the electronic properties of adjacent bilayer regions. Therefore, these smaller gapped regions play a more important role in determining the electronic properties of the gapped bilayer graphene, which may account for the subgap conductance observed in Bernal bilayers[7].

In summary, we show that there are two types of domain walls in gapped graphene bilayers and we also demonstrate that the domain walls separating two different bilayer regions will show novel electronic properties not present in the adjacent graphene bilayer regions. Our observation of exotic Landau quantization in the domain walls opens up exciting opportunities to explore novel electronic properties and topological phases in bilayer graphene by using STM and STS.

**Acknowledgments**

This work was supported by the National Basic Research Program of China (Grants

Nos. 2014CB920903, 2013CBA01603), the National Natural Science Foundation of China (Grant Nos. 11422430, 11374035), the program for New Century Excellent Talents in University of the Ministry of Education of China (Grant No. NCET-13-0054), Beijing Higher Education Young Elite Teacher Project (Grant No. YETP0238). L.H. also acknowledges support from the National Program for Support of Top-notch Young Professionals.

**Figure Legends:**

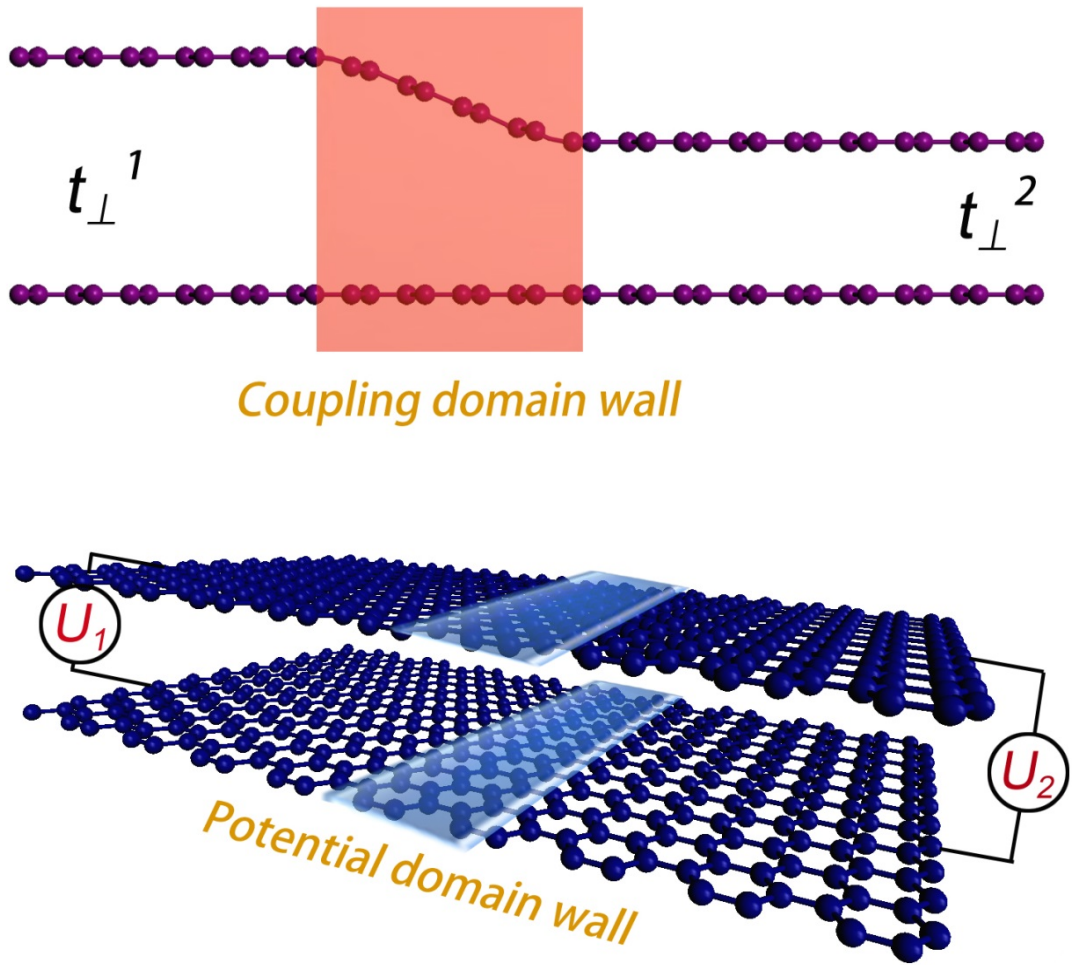

**Figure 1 |** Illustrations of two types of domain walls in graphene bilayers. Upper plane: a coupling domain wall separates two regions with different nearest-neighbour interlayer hopping parameters, which arise from different interlayer distances. Lower plane: a potential domain wall divides two bilayer regions with different chemical potentials. The chemical potential difference between the two layers can be introduced by applying a vertical electrical field or by a disorder potential induced by the supporting substrate.

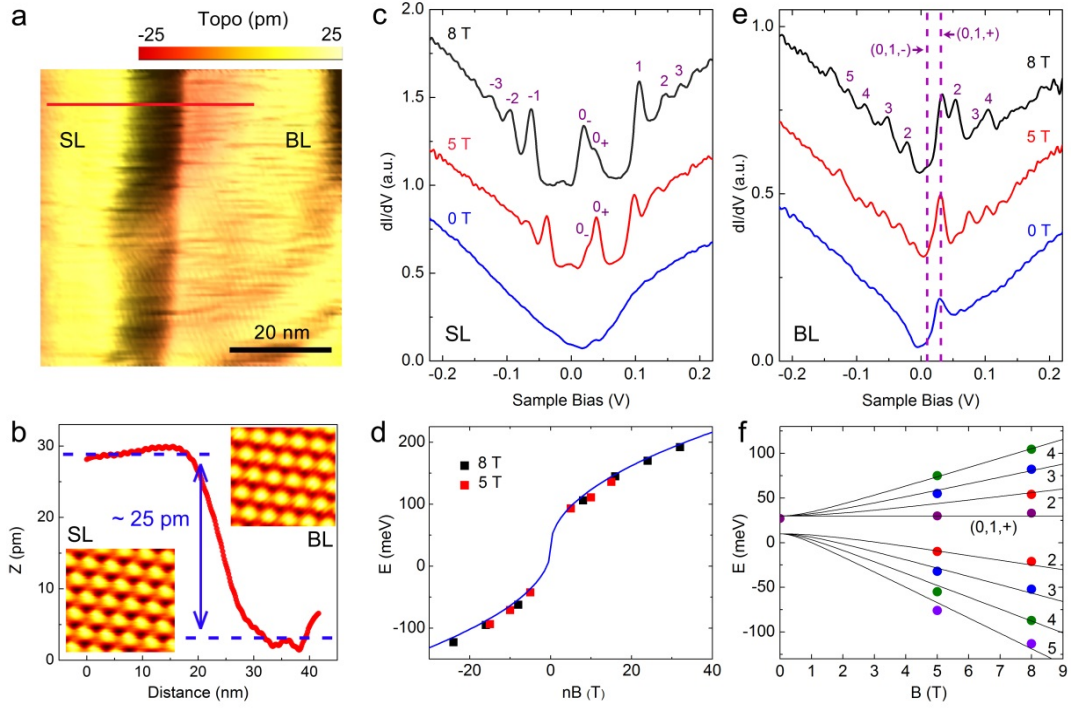

**Figure 2 |** STM image around a coupling domain wall and Landau quantization in monolayer and bilayer regions. **a**. A STM image of a coupling domain wall separating single-layer (SL) and bilayer (BL) grapheme regions. **b**. The height profile along the red line in **a**, indicating the interlayer distance of the left side is slightly enlarged. The insets show atomic resolution images of the SL and BL regions. The triangular contrasting recorded in the SL is attributed to the broken symmetry of sublattices. **c** and **e** show tunnelling spectra recorded under various magnetic fields in the SL and BL regions, respectively. For clarity, the curves are offset in Y-axis and LL indices are marked (here +, - are valley indices). **d**. The LL peaks energies extracted from **c** plotted against (nB). The solid curve is the fit of the data with Eq. (2), yielding a Fermi velocity $v_F = (0.82 \pm 0.01) \times 10^6$ m/s. **f**. The LL peaks energies obtained from **e** plotted against *B*. The solid curves are the fit of the data with Eq. (3).

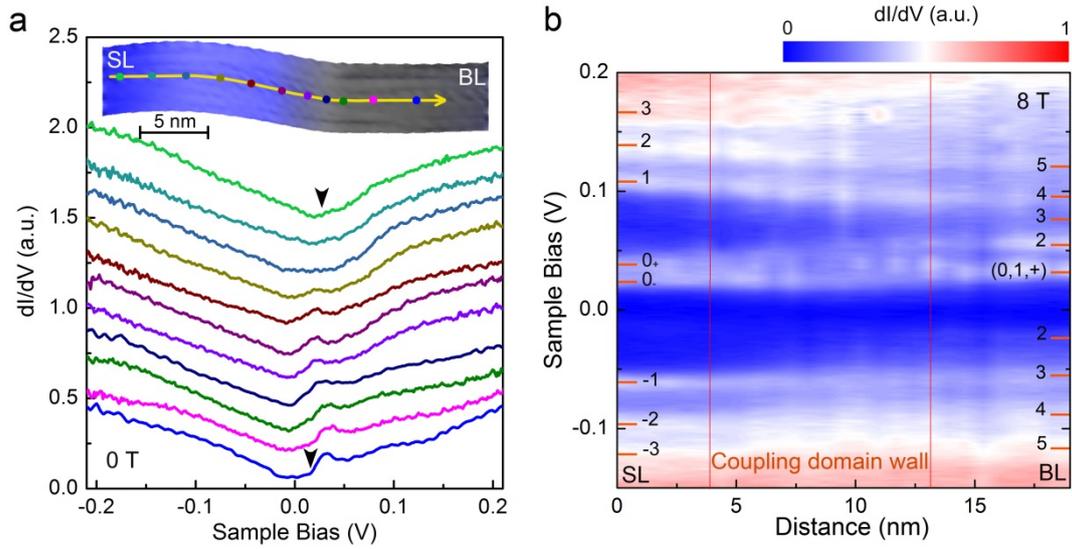

**Figure 3 |** Evolution of tunnelling spectra across the coupling domain wall. **a**. Zero-field spectra taken at different positions marked in the STM image of the inset. The charge neutral point (as marked by black arrows) located at ~ 30 meV in the SL region, and shift to ~ 20 meV in the BL region. **b**. Evolution of the Landau levels across the coupling domain wall. The colour scale encodes the magnitude of the differential conductance and the LL indices in the SL and BL regions are marked. From the LL map, the width of the domain wall is estimated to be about 10 nm.

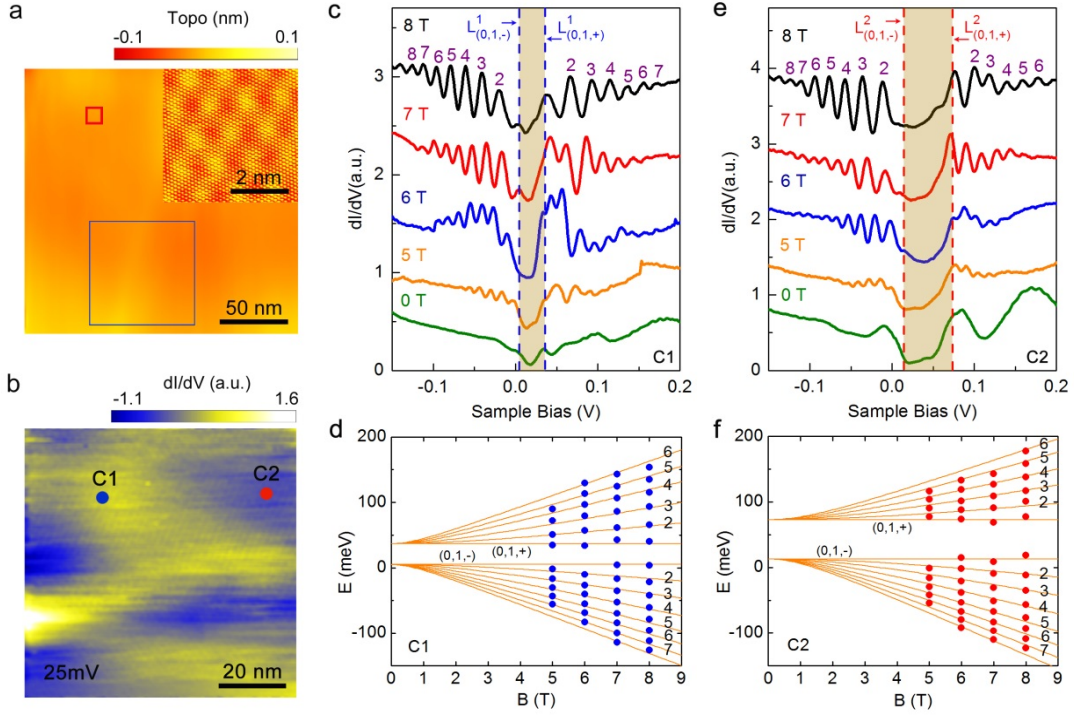

**Figure 4 | STM image and Landau quantization around a potential domain wall. a**. A 200 nm × 200 nm STM topographic image of a decoupled graphene bilayer with a potential domain wall. Inset: Zoom-in STM image of the red frame shows moiré pattern structure. The atomic-resolution image shows triangular lattice, indicating that the topmost two graphene sheets are Bernal bilayer. The period of the moiré pattern is $D = (2.26 \pm 0.05)$ nm, which is induced by a twisted angle $\theta = (6.2 \pm 0.1)°$ between the second graphene sheet and the third sheet. **b**. $dI/dV$ map with a fixed sample bias of 25 mV recorded in the blue frame in panel **a**. **c** and **e** show STS spectra taken at positions C1 and C2 in panel **b**, respectively. The band gaps around the positions C1 and C2, *i.e.*, the energy spacing between $LL^1_{(0,1,-)}$ ($LL^2_{(0,1,-)}$) and $LL^1_{(0,1,+)}$ ($LL^2_{(0,1,+)}$), and the LL indices are marked. **d** and **f** show LLs peak energies obtained in panel **c** and **e** plotted against magnetic field *B*, respectively. The solid curves are the fitting result with Eq. (3). The charge neutral points around the positions C1 and C2 are estimated to locate at 21 meV and 43 meV, respectively.

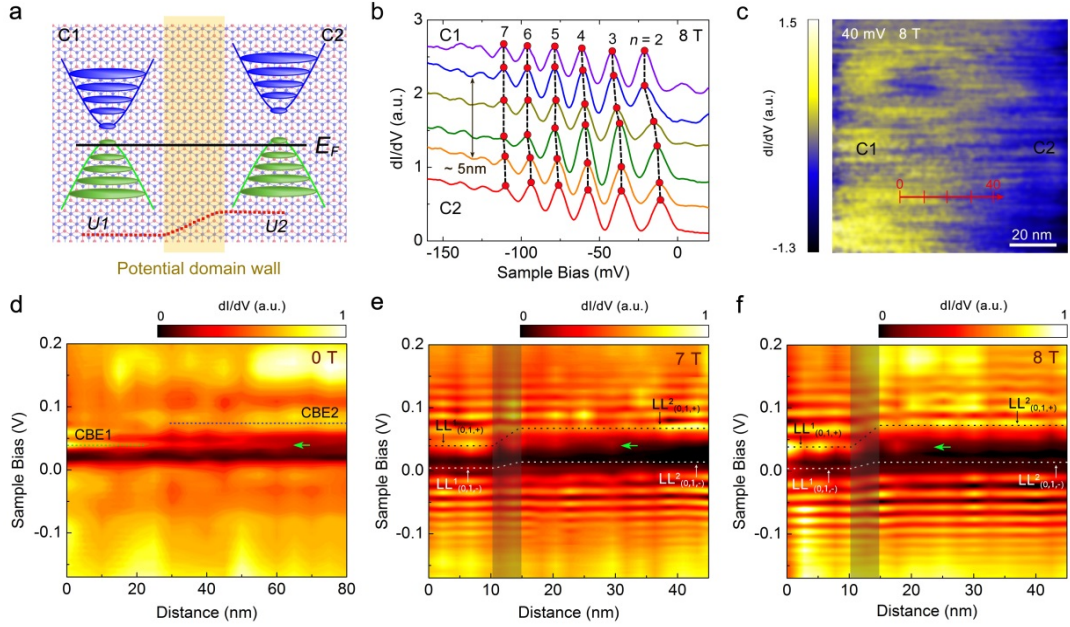

**Figure 5 |** Evolution of tunnelling spectra across the potential domain wall. **a**. Schematic image of the potential domain wall in the experiment. **b**. Spatial evolution of the LLs below the charge neutrality point across the potential domain wall. **c**. High field, 8 T, *dI/dV* map recorded around the potential domain wall at a fixed sample bias of 40 mV. **d-f**. Evolution of the spectra recorded at different magnetic fields across the potential domain wall. The domain wall is marked by the shadow regions in panels **e** and **f**. The green arrow in panel **d** marks the farthest position where we can detect the signal of CBE1. The green arrows in panels **e** and **f** mark the farthest positions where we can detect the signal of $LL^1_{(0,1,+)}$ in the magnetic fields of 7 T and 8 T, respectively.